\documentstyle[amssymb,preprint,aps]{revtex}

\begin{document}

\title{TOWARDS MICROSCOPIC THEORY OF PHASE TRANSITIONS: CORRELATION RADII
AND CRITICAL INDICES}
\author{Mark E. Perel'man}
\address{Racah Institute of Physics, Hebrew University, Jerusalem, Israel}
\date{\today}
\maketitle

\begin{abstract}

From the microscopic point of view almost all bonds between
particles of condensed substances must be performed by exchanges
of virtual photons. Consequently the duration of their virtuality
must be longer than the extent of their free path in the
substance, the magnitudes of all expressions in such inequality
are known from low frequencies scattering. This approach allows to
suggest that the break of some set of bonds of particles, i.e. the
phase transitions, will be originated just at the reversing of
established inequality. Such assumption leads to definition of the
radius of correlations or bonds: $R_{c}\symbol{126}\ E^{-2/3}$
that proves the universality of this critical index. The energies
E, which can be liberated at phase transitions, are definite for
different types of critical phenomena. Reformulation of the
Ginzburg-Landau model of phase transitions via expansion of
thermodynamic potentials over $R_{c}$, instead temperatures
distance, leads to the correct system of all critical indices.

\end{abstract}


\qquad \qquad

{\large INTRODUCTION}\bigskip

All common phenomenological theories of phase transitions and critical
phenomena are formulated via the order parameters (recent reviews [1 - 3]).
Therefore the simplest way for approach to a microscopic theory must
imperatively include microscopic analysis and deduction of suitable
parameters.

Microscopic theories of phase transitions must be based on consideration of
elementary processes of interaction between particles of condensate. So the
Van-der-Waals and Casimir interactions are represented via exchange of
virtual photons, i.e. in the framework of quantum electrodynamics (QED) [4,
5]. These representations are needed, at least, for justification of results
obtained by phenomenological methods.

All thus considered bonds are performed by exchange of virtual photons. But
with increasing of the free flight path between interactions a possibility
of converting of virtual photon into real one is growing. In this way can be
suggested that just at approach to critical points these photons become
turning into the real state and, in particular, could be emitted, i.e. they
are on the verge of completeness of their formation process, their
''dressing''.

In such direction we {\it propose} that the problem of phase transitions can
be expressed in terms of internal fields. It will be shown that this way
elucidates the physical sense of the basic phenomenological hypotheses. In
this way the consideration of singularities of thermodynamical magnitudes
has a heuristic, at least, sense, since allows the simple verification of
our suggestion ({\it virtual photon - real photon transitions}) by
demonstration of some its prospects in understanding of certain complicated
problems.

For these aims some temporal magnitudes must be briefly considered.

The first of them is the function of duration of time delay at elastic
scattering introduced by Wigner as the derivative of partial phase shifts, $%
\tau _{l}(\omega )=d\delta _{l}/d\omega $ [6], generalized by Smith [7] via
the $S$-matrix of scattering:

$\tau _{1}=$ Re $(\partial /i\partial \omega )\ln S$. \qquad \qquad (1.1)

This definition can be expressed through the response functions also [8].

The function of duration of final state formation was introduced by Pollak
and Miller [9] and can be expressed as

$\tau _{2}=$ Im $(\partial /i\partial \omega )\ln S=(\partial /\partial
\omega )\ln |S|$. \qquad \qquad (1.2)

Such expressions usually are introduced into theory artificially, ad hoc
(e.g. the reviews [10]). But at the QED calculations of multiphoton
processes, where the possibility of subsequent absorption of photons is the
crucial factor, they are arise automatically [11, 12].

Here must be underlined that the expression, similar to (1.1), had been
appeared, well before the Wigner delay time, in the Uhlenbeck and Beth
expression of second virial coefficient of molecular interactions [13].
Their approach (e.g. [14, 15]) leads to such part of the second virial
coefficient describing pair interactions (here and below $\kappa _{B}=1$):

$\Delta B(T)\ \symbol{126}-N%
\mathop{\displaystyle\sum}%
_{l}(2l+1)%
\displaystyle\int %
d{\bf k}\exp (-\hslash ^{2}k^{2}/2mT)\ d\delta _{l}/dk$.\qquad \qquad\ (1.3)

The derivative of phase shift can be rewritten via energy and this
expression can be evidently treated in temporal terms [16]. Hence the
density of interacting particles is proportional to the duration of their
interactions, averaged over the Boltzmann distribution, and it shows, that
the temporal approach has deep roots and perspectives in statistical
mechanics. The possibilities of $S$-matrix formulation of statistical
mechanics [17], where originated the expressions of (1.1) type, can be of
great concern in such approach.

In the papers [18] are shown that both definitions (1.1) and (1.2) can be
formally joined and combined as $\tau (\omega ,{\bf r})\equiv \tau
_{1}+i\tau _{2}=(\partial /i\partial \omega )\ln S(\omega ,{\bf r})$. This
expression is tantamount to the equation

$(\partial /i\partial \omega )\ S(\omega ,{\bf r})=\tau (\omega ,{\bf r})\
S(\omega ,{\bf r})$,\qquad \qquad\ (1.4)

which determines temporal functions through the response functions or vice
versa. Formally this expression is presented as the reciprocal form of
Schr\~{o}dinger equation for $S$-matrix, rewritten via some Legendre-type
transformation $x_{\mu }\longleftrightarrow p_{\mu }$ and with a
''temporal'' operator $\tau (\omega ,{\bf r})$ instead the Hamiltonian. Note
that a similar equation can be considered as some transformation of the
Bloch equation with $\beta \equiv 1/T\rightarrow i\omega $, i.e. leads to
the immediate consideration of temporal functions in quantum statistics. In
such equations response functions, matrix elements of processes or even
density matrices can be considered instead of $S$.

The temporal equation (1.4) can be established, in particular, in the scope
of the formal theory of scattering and temporal functions can be formally
written as the propagators of particles [12, 18], that allows to express
thermodynamic potentials through the temporal functions (Section 2). This
theorem for the grand potential shows that the density of interacting
particles linearly depends on the time duration of interaction and therefore
predetermines the dependence of the ''volume of interaction'' on temporal
magnitudes. By this theorem, in particular, the relation of latent heat to
temperature of transition can be estimated (the Appendix).

The volume of interactions of virtual photons with bound electrons can be
introduced as the product of total cross-section and duration or path of
completion of their interaction (Section 3). Their dependence on energy
allows to express the volume of interaction and, consequently, the radius of
interaction, i.e. the correlation radius $R_{c}$, through the energy of
transition, its temperature or other suitable parameters' ''distance'' till
critical point. Hence the universality of critical indices and conformance
of their values with the known experimental data are shown in the Section 4.

In the Section 5 the some refinement of the Ginzburg-Landau theory of phase
transition [19] is considered. Instead of the magnitudes of reduced
temperature $\epsilon =1-T/T_{c}$, chosen usually as the simplest
non-dimensional parameter, the expansion of thermodynamic potentials over
the values of $R_{c}N^{1/3}$ is performed ($N$ is the density of
scatterers). It leads to modifications of criteria of existence of several
states and allows experimental verifications of suggested theory.

The Section 6 contains remarks about some peculiarities of $2$-$D$ phase
transitions. This problem can be significant, in particular, at
consideration of layered media. Some further perspectives are considered in
the Conclusions.

Notice that the first attempts in the suggested direction were undertaken in
[20] at consideration of liberation of latent heat at phase transitions by
its converting into characteristic radiation. These results were
strengthened by another examinations [21] and by some supporting experiments
[22]. To this direction can be referring also the microscopic description of
optical dispersion via durations of elementary acts of electron-photon
unbiased elastic scatterings [23].

\

{\large 2. TEMPORAL FUNCTIONS AND THERMODYNAMIC POTENTIALS}

Let express thermodynamic potentials in terms of temporal functions.

We recall from nonrelativistic statistical mechanics (e.g. [14]) that the
grand potential $\Omega $ is given by

$\Omega =-\frac{1}{\beta }\ln \ $tr$\ \exp \{-\beta ({\bf H}-\sum_{k}\mu
_{k}N_{k})$\}$\rightarrow -\frac{1}{\beta }\ln \sum N\{e^{-\beta \mu N}\ $tr$%
_{N}\ e^{-\beta {\bf H}}\}$ \qquad (2.1)

with the Hamiltonian ${\bf H}$, $\beta =1/T$ and chemical potentials $\mu
_{k}$.

For simplicity we assume that ${\bf H}={\bf H}_{0}+{\bf V}$ and express
(2.1) via propagators of interacted and free particles (resolvents), $%
G(E)=(E-{\bf H})^{-1}$, $g(E)=(E-H_{0})^{-1}$, as functions of complex
energy $E\rightarrow E+i\Gamma $, $\Gamma \geq 0$ in passive media. With
this aim we shall take into account the representation:

tr $e^{-\beta {\bf H}}=\frac{-1}{2\pi i}%
\displaystyle\oint %
dEe^{-\beta E}$ tr $G(E)\rightarrow \frac{-1}{\pi }%
\displaystyle\int %
dEe^{-\beta E}$ Im tr$\{G(E)-g(E)\}$, \qquad (2.2)

where the counterclockwise counter in first integral extends over the
spectrum of $H$ and the last expression contains the connected diagrams only.

The main factor of last expression can be written in the terms of $S$-matrix:

Im tr$\{G(E)-g(E)\}$ = $\frac{1}{4i}$ tr$\{(\partial
_{E}S^{-1})S-S^{-1}\partial _{E}S\}$. \qquad \qquad (2.3)

The comparison with (1.1) shows, that this expression is proportional to the
time duration (delay) of scattering process. In a slightly another manner,
as was found in [12, 18], the complex time duration of interaction, i.e. the
difference between durations of particles flight with and without
interaction, can be expressed as

$\tau (\omega ,{\bf r})\equiv \tau _{1}+i\tau _{2}=i(G(E)-g(E))$. \qquad
\qquad (2.4)

Hence the grand potential can be written as

$\Omega =(\pi \beta )^{-1}\ln
\mathop{\displaystyle\sum}%
_{N}\left\{ e^{-\beta \mu _{N}}%
\displaystyle\int %
dEe^{-\beta E}tr_{N}(\tau _{1})\right\} $,\qquad \qquad\ (2.5)

which shows possibility of presentation of all thermodynamic functions via
temporal parameters. The relations of Kramers-Kr\"{o}nig type interconnect
the functions $\tau _{1}$ and $\tau _{2}$ [18], and a jump or sharp changing
of $\Omega $ must correspond to changing of both $\tau $ functions and vice
versa. Thereby we propose that phase transitions may be described
microscopically via consideration of changing of $\tau $ functions, i.e. via
variation of durations of particles interactions and formations.

Let's consider such simple model for the first kind phase transitions [20]:
system of different phases correspond to a system of levels, distances
between them are of order of latent heat of transition per particle. So, for
the system of two levels-phases $E_{0}=\hbar \omega _{0}$, upper level has a
width $\hbar \Gamma $. Such model suggests the temporal function for
particles interaction:

$\tau (E)=\frac{1}{i(\omega \ -\ \omega _{0})\ +\ \Gamma /2}\equiv \frac{1}{%
i(\omega \ -\ \varepsilon )}$ ,\qquad \qquad\ (2.6)

It can be supposed several hypothetical forms of an width $\Gamma $, but for
our consideration only the natural assumption $\Gamma <<\omega _{0}$ will be
essential. (Notice, that this condition can be non-correct in the nearest
vicinity of critical points!)

Similar forms of temporal functions can be connected with redistribution of
internal energy between degrees of freedom and so on. Thereby unlike the
first order transitions this energy is not emitted, but the expressions of
(2.6) type will be assumed below for all considered types of transitions,
without further discussions, as the simplest ones.

The expression (2.6) must be averaged over the Boltzmann distribution, the
time interval of averaging $\Delta T=(\hbar /E_{1})$ contains a big, but
non-concretized energy $E_{1}$. It leads to such expression and its
semi-regular expansion [24]:

$\overline{\tau }(E)/\Delta T=ie^{-\beta \varepsilon }$Ei($\beta \varepsilon
$) $\rightarrow i%
\mathop{\displaystyle\sum}%
_{1}(-1)^{n+1}n!(\beta \varepsilon )^{-n}$.\qquad \qquad\ (2.7)

For our aims the first terms of (2.7) are sufficient:

$\overline{\tau }_{1}(E)\ \symbol{126}\ \varphi (T)\ \hbar \Gamma /2(\omega
_{0}^{2}+\Gamma ^{2}/4)$, \qquad \qquad (2.8)

$\overline{\tau }_{2}(E)\ \symbol{126}\ \varphi (T)\ \hbar \omega
_{0}/2(\omega _{0}^{2}+\Gamma ^{2}/4)$, \qquad \qquad (2.9)

where $\varphi (T)$ $=\Delta T/T$ is a slow varying function of temperature.
Moreover for all following consideration will be essential only such
propositions:

$\overline{\tau }_{2}(E)$ $\symbol{126}\ 1/\omega _{0}$ and $\overline{\tau }%
_{2}(E)$ $>>\overline{\tau }_{1}(E)\ \symbol{126}\ (\Gamma /\omega _{0})%
\overline{\tau }_{2}$. \qquad \qquad (2.10)

Note that an estimation of $\overline{\tau }_{2}$ corresponds, in the
according with (1.2), to the usual asymptotic representation of Green
function in microscopic theories of phase transitions (e.g. [25, 26 ]) as $%
G(p)\ \symbol{126}\ Z\ p^{-2+\eta }$ with $\eta \longmapsto +0$. For more
scrupulous considerations of temporal functions their correspondence with
propagators, $\tau (\omega ,{\bf r})=iG(\omega ,{\bf r})$, can be used: it
follows from the formal theory of scattering with $G(p)\ \symbol{126}\ ({\bf %
H}-E)^{-1}$. The reciprocal Schr\"{o}dinger equation (1.4) can be compared
with the Ward-Takahashi identity of QED for the derivative of Green
functions, and it leads to more complete expressions for temporal functions.
In statistical physics the analogical role must be performed by the
Landau-Pitaevskii identity [27]:

$(\partial /\partial p_{0})G(p)=-(2\pi )^{-4}\{G^{2}(p)\}^{\omega }[1-i\int
(\Gamma (p,q))^{\omega }\{G^{2}(p)\}^{\omega }]d^{4}q$,\qquad\ (2.11)

which is written in the covariant notation and where are omitted, for
simplicity, the tensor indices, $\{G^{2}(p)\}^{\omega }=%
\mathrel{\mathop{\lim }\limits_{k\rightarrow 0}}%
G(p)\ G(p+k)$ and analogical definition for the vertex operator $\Gamma
_{\beta \delta ,\alpha \delta }(p,q)$. From (2.11) follows that

$\tau (p)=iG(p)+G(p)%
\displaystyle\int %
G(q)\ \Gamma (p,q)\ G(q)\ dq$, \qquad \qquad (2.11')

which allows improving of temporal functions, etc. But for an estimation of
critical indices, as the most demonstrative presentation of the theory, in
this paper it would be sufficient to restrict the consideration by the
simplest form (2.10).

\qquad

{\large 3. SCATTERING AMPLITUDES AND VOLUMES OF CORRELATIONS}

Let's discuss the peculiarities of (virtual) photons exchange that must
predetermine the main properties of unbounded condensed media in the QED
representation.

For this aim the comparison of two temporal magnitudes, the duration of
formation of ''physical'' photons $\tau _{2}$ and the time of their flight
on distance of mean free path in a substance $\ell /c$, is needed. In this
case $\ell =1/\sigma _{tot}N$, where $\sigma _{tot}(\omega )$ is the total
cross-section of $\allowbreak e$-$\gamma $ interactions and $N$ is the
density of scatterers (outer electrons). Such description is consistent with
the theory developed in [23], where it is shown that propagation of photons
in media can be considered as free flights with the vacuum speed $c$ between
delays on $\tau _{1}$ at scatterers.

If for some frequencies $\tau _{2}(\omega )>\ell (\omega )/c$, then {\it it
can be suggested} that photons of corresponding frequencies, performing
exchange bonds between particles of substance, are remained virtual. The
zone of space extent of their virtuality corresponds to a near field of
classical electrodynamics, i.e. is of order of $c\tau _{2}$. At the point of
reversing of this inequality, as we {\it assume}, the bonds will be ruptured
and phase transitions will be originated.

Such proposition, expressed as $\tau _{2}\geq \ell /c$, leads to the basic
inequality, that can represent the definition of condensed substance as the
state, in which constituents interact via exchange of virtual photons of
definite frequencies:

$c\tau _{2}\sigma _{tot}N\equiv V_{c}(\omega )N\geq 1$.\qquad \qquad\ (3.1)

The magnitude $V_{c}(\omega )$ may be named ''the effective volume of
electromagnetic interactions on the frequency $\omega $'' or more shortly
''the volume of interaction''; its inverse value $1/V_{c}(\omega )$
represents the density of interacting particles in agreement with (2.4).
This magnitude can be generalized by taken into account the space-time
dependencies, anisotropy of states, polarization effects, external fields,
etc. The corresponding frequency corresponds to the potential energy and
must be connected with the latent heat, energy of atomization and so on.
Notice, that at the low frequency limit $c\tau _{2}\approx c/\omega $ , this
value is twice bigger than the value of uncertainty and therefore is
measurable. The non-dimensional magnitude $X=N\ V_{c}(\omega )$ or its
suitable degree would be considered as the order parameter. Some other
descriptions of this main magnitude will be given below.

The proposition (3.1) leads to the introducing of effective radius of
electromagnetic (EM) correlations or the length of EM interactions on the
frequency $\omega $ as

$R_{c}(\omega )=(3c\tau _{2}\sigma _{tot}/4\pi )^{1/3}$ \ \ \qquad (3.2)

or nondimensional parameter

$\eta $ = $R_{c}(\omega )\ N^{1/3}$\qquad \qquad\ (3.2')

and conditions for the saturation of interactions of definite type on this
frequency as

$V_{c}(\omega )\ N=1$ or $\eta =(3/4\pi )^{1/3}$, \qquad \qquad (3.3)

Note that this condition can be expressed via the plasma frequency $\omega
_{P}=(4\pi e^{2}N/m)^{1/2}$, that allows several interpretations for
long-range interactions.

The induction of external influence upon medium can be defined as

$F(\omega )=%
\mathop{\displaystyle\sum}%
_{1}^{\infty }(V_{c}N)^{n}\ I(\omega )=\frac{V_{c}N}{1\ -\ V_{c}N}I(\omega
)\equiv \chi (\omega )\ I(\omega )$, \qquad \qquad (3.4)

where $\chi (\omega )$ is the generalized susceptibility of medium.
Therefore the proposed conditions for phase transitions can be determined as
singularities of $\chi (\omega )$ and/or of $R_{c}(\omega )$ at the definite
frequencies.

In the general case of $3$-D space interactions the total cross-section is
expressed by the optical theorem of QED as

$\sigma _{tot}=(4\pi c/\omega )$\ Im $A(0)$, \qquad \qquad (3.5)

where $A(0)$ is the amplitude of elastic scattering on the zero angle. For
the processes of photons exchanging at low frequencies Im $A(0)\rightarrow
r_{0}=e^{2}/mc^{2}$. It leads to the volume of EM correlations

$V_{\gamma }=4\pi e^{2}/m\omega ^{2}\equiv 4\pi r_{0}\ (\hbar c/\hbar \omega
)^{2}$ \qquad \qquad (3.6)

and to the radius of such correlations:

$R_{\gamma }=(3V_{\gamma }/4\pi )^{1/3}=(3c^{2}r_{0})^{1/3}\omega
^{-2/3}=0.28\times 10^{-4}$ $\lambda ^{2/3}$ [cm]. \qquad \qquad (3.7)

Notice, that the volume of interaction can be determined also via the
maximal cross-section $\sigma _{\max }$ $\symbol{126}\ \lambda ^{2}$ and the
minimal duration of final state formation $\tau _{\min }=r_{0}/c$, it leads
to the same results. The strange, at the first sight, result: $V_{\gamma }\
\symbol{126}\ k^{-2}$ and therefore $V_{\gamma }\ \symbol{126}\ R^{2}$ would
be foreseen from the well-known dependence of the Fermi energy: $E_{F}\
\symbol{126}\ (N/V)^{2/3}\ \symbol{126}\ R^{-2}$.

Another interpretation of the magnitude of interaction volume comes from the
Fourier transform of expression (3.6):

$V_{c}({\bf r})\equiv \frac{1}{(2\pi )^{3}}%
\displaystyle\int %
d{\bf k}V_{c}({\bf k})e^{i{\bf kr}}=\frac{r_{0}}{r}\equiv \frac{e^{2}/r}{%
mc^{2}}$, \qquad \qquad (3.8)

i.e. it represents relation of the Coulomb energy to the rest mass of
interacting particles and can be stated as the basic principle for all
consequent consideration.

At $r\rightarrow a_{B}$, the Bohr radius, this expression leads to the
doubled Rydberg constant. For the usual distances of order $R_{0}$ $\symbol{%
126}10^{-8}$ cm between atoms/molecules in condensed state the estimation
(3.7) leads to $\lambda \ \symbol{126}\ 6.8\times 10^{-6}$ cm, the
reasonable order of magnitude for a long-range, at least, interatomic
correlations.

Apart from this general bindings of atoms/molecules there must be considered
additional interactions dictated by the specific parameters of substance
constituents. It can be proposed that these interactions would lead to the
specific volumes of interactions, providing the peculiarities of these
substances. Let's briefly consider some of them.

At processes of virtual photons exchange between particles with magnetic
moment $\mu $ or between spherical rotators with dipole moment $d$ and
moment of inertia $I$ the corresponding amplitudes are:

$A_{m}=4\mu ^{2}\omega /c$; \qquad \qquad (3.9)

$A_{d}=(4\omega ^{2}d^{2}/3Ic^{2})/|\hbar ^{2}-\omega ^{2}I^{2}|$. \qquad
\qquad (3.10)

In substances, where the electrons exchange forces must play the significant
role, the corresponding matrix element is of order

$A_{ee}=e^{2}/mv^{2}\rightarrow \ e^{2}/3T$ \qquad \qquad (3.11)

and this can lead to long-distance correlations. But \$A\_\{%
\mbox{$\backslash$}%
gamma \}\$ far exceeds these amplitudes, and since all these interactions do
not exclude the direct virtual photons exchange, the effective radius for
such properties of these substances as solidification or condensation
remains equal to (3.7). It seems that all other interactions are related to
the long-range correlations only and correspond, in essence, to the second
order transitions.

The values of both temporal parameters must be measurable. For this aim can
be considered the group index of refraction [23] expressed via the QED
parameters of elastic $e$-$\gamma $ scattering as

$n_{gr}(\omega )\equiv cdk/d\omega =1+cN\sigma _{tot}\tau _{1}$. \qquad
\qquad (3.12)

The ''group'' coefficient of extinction must be correspondingly expressed as

$\kappa _{gr}(\omega )=cN\sigma _{tot}\tau _{2}$\qquad \qquad\ (3.12')

that coincides with $V_{c}N$. And since these ''group'' indices are
determined via the usual complex index of refraction as $n_{gr}+\kappa
_{gr}=(d/d\omega )[\omega (n+i\kappa )]$ and $\kappa =(c/2\omega )\gamma $,
where $\gamma (\omega )$ is the linear coefficient of absorption, the volume
of interaction is expressed via the measurable magnitudes:

$V_{c}(\omega )=2N^{-1}(d/cd\omega )\ \gamma (\omega )$. \qquad \qquad (3.13)

Hence the introduced parameter can really map the peculiarities of
interatomic bonds.

As must be specially underlined, just such singularities of (3.12) conforms
to the residual rays (dielectric anomalies) at reflection on perfect crystal
surfaces, which are experimentally known for a very long time and are
empirically, without whatever theoretical justifications, connected to the
latent heat of melting [28, 29]. The suggested approach leads to
consideration of this effect as the direct corroboration of existence of
resonant emission of latent heat at phase transitions.

\qquad

{\large 4. RADII OF CORRELATIONS AND UNIVERSALITY OF CRITICAL INDICES}

As the volumes of interaction are expressed via frequencies of interactions,
they must be connected now with macroscopic parameters, at least closely to
critical points.

The mole latent energy of phase transitions of the first kind $\Lambda
=T(S_{2}-S_{1})$, where $S_{k}$ is the entropy of $k$-th phase. If at
approach to the critical temperature $T_{c}$ this difference smoothly
aspires to zero, the latent energy per atom/molecule can be presented as

$W\equiv \Lambda /N_{A}\approx T\ (\partial S/\partial T)\
(T_{c}-T)\rightarrow \ \epsilon \ T_{c}^{2}\ (\partial S/\partial T)_{T=Tc}$
\qquad \qquad (4.1)

with $\epsilon =|T_{c}-T|/T_{c}$.

If at transition into a lower energy state the latent heat $\hbar \omega $
is liberated by emission of single quantum, the parity conservation law
requires the changing of symmetry of transient particles. Therefore the
transitions with emission of two photons should be the most preferable
(notice that just the two-photon exchanges lead to the Van-der-Waals bonds
[4]). If these two photons will be emitted as $^{1}S_{0}$ system per each
transient particle, it will be equivalent to emission of scalar boson, and
just such possibility corresponds to condition for spontaneous breaking of
symmetry (the Goldstone theorem, e.g. [30]), needed at crystallization or
other transitions into lower symmetry states.

Let's suppose, for simplicity, that this energy is distributed between $n$
bonding quanta of equal frequencies in (3.6): $\hbar \omega \rightarrow W/n$%
. It leads to the volume of interaction:

$V_{c}(W/n)=V\prime \epsilon ^{-2}$ \qquad \qquad (4.2)

with $V\prime \approx $ const, and to the radius of virtual photon's
correlations:

$R_{c}(T)=(3V_{c}/4\pi )^{1/3}=R_{0}\ \epsilon ^{-2/3}$. \qquad \qquad (4.3)

For the phase transitions of second kind the thermal energy of single
particle at close range to the critical temperature is determined as

$W=\frac{1}{N_{A}}%
\displaystyle\int %
_{T}^{T_{c}}CdT\approx (\overline{C}T_{c}/N_{A})\ \epsilon $, \qquad \qquad
(4.1')

i.e. with the similar functional form of radius of correlations at $%
W\rightarrow \ n\hbar \omega $.

Analogical forms may be determined for other types of interactions: for the
electrical and magnetic dipole interactions, correspondingly, $\hbar \omega
\rightarrow \ {\bf dE}$ or\ $\mu {\bf H}$, etc. It naturally leads to the
similar radii of correlations:

$R_{c,d}=R_{0,d}\ (|{\bf dE}|/|{\bf dE}|_{c,d})^{-2/3}$;\qquad \qquad\ (4.4)

$R_{c,m}=R_{0,m}\ (|\mu {\bf H}|/|\mu {\bf H}|_{c,m})^{-2/3}\qquad \qquad $
(4.4')

and analogical for higher moments.

The electron exchanging, connected with (3.11), leads to the correlation
radius:

$R_{c,T}=R_{0,T}\ (T/T_{c,e})^{-2/3}$.\qquad \qquad\ (4.5)

Hence, in all these cases the critical index of correlations is of the
universal type with $\eta $ or $R_{c}$ as

$R_{c}\symbol{126}$ $\epsilon ^{-\nu }$, $\qquad \nu =2/3$, \qquad \qquad
(4.6)

where $\epsilon \ \symbol{126}\ \{|T_{c}-T|,|{\bf dE}|,|\mu {\bf H}%
|,T,\ldots \}$ for different types of critical phenomena. In the scope of
proposed theory this conclusion supports the hypothesis of transitions
similarity, i.e. the universality of critical behavior.

The dependence (4.6) predetermines estimations of other critical indices.
Thus, by the Ornstein-Zernike pair correlation function in the unit volume,
represented as

$G({\bf r})\ \symbol{126}\ (1/R_{0}^{2}r)\exp (-r/R_{c})$, \qquad \qquad
(4.7)

the averaged square of fluctuations of the order parameter is expressed as

$\left\langle (\Delta \eta )^{2}\right\rangle =%
\displaystyle\int %
G(r)dV\ \symbol{126}\ 4\pi (R_{c}/R_{0})^{2}$. \qquad \qquad (4.8)

It means that the critical behavior of susceptibilities is universally
determined as

$\chi =(V_{0}/T)\left\langle (\Delta \eta )^{2}\right\rangle \ \symbol{126}\
\epsilon ^{-\gamma }$, $\qquad \gamma =4/3$. \qquad \qquad (4.9)

Just this index determines, for example, the critical behavior of isothermal
compressibility, the intensity of scattering at the phenomenon of critical
opalescence:

$I(q)/I_{0}(q)=\frac{1}{N}%
\displaystyle\int %
G(r)e^{-i{\bf qr}}dV=1/\pi R_{0}^{2}(q^{2}+R_{c}^{-2})$ \qquad \qquad (4.10)

with the transfer moment $q\rightarrow 0$ and so on.

Notice, that the analogical phenomena of critical opalescence should be
observable also at ${\bf H}\rightarrow 0$ or ${\bf E}\rightarrow 0$ in
transparent ferromagnetics, ferroelectrics and liquid crystals.

The definition of volume of interaction allows the assessing of some other
physical quantities. So, for example, the maximal temperature for
possibility of existence of chemical bonds can be estimated. Let's define
for this aim the $e$-$e$ interaction volume (3.11) averaged over the Maxwell
distribution: such averaging implies the substitution $(\hbar /p)\rightarrow
\hbar \ /(2\pi mT)^{1/2}$ and therefore the volume of interaction, written
with taking into account of (4.5), can be expressed as

$V_{ee}=4\pi (\hbar /p)^{2}A_{ee}\rightarrow \left\langle V\right\rangle
_{T}=\frac{2}{3}r_{0}(c\tau _{T})^{2}$, \qquad \qquad (4.11)

where $\tau _{T}=\hbar /T$. It shows that the electron exchange interactions
at the distances $R>10^{-8}$ cm are possible only at $T<T_{\max }\symbol{126}%
\ 10^{4}K$.

As was mentioned, it is assumed that substances would be in a condensed
state if and only if the free path length of internal photons is not bigger
then the correlation radius: $\ell \leq \ R_{c}$. This inequality can be
rewritten as

$\hbar \omega _{\max }\leq \hbar c[3(4\pi )^{3}N^{3}r_{0}^{4}]^{1/5}$,
\qquad \qquad (4.12)

which for the substances density $N$ $=10^{21}\div 10^{24}$ particles/cm$%
^{3} $ leads to the bound energy of order $0.04\div 2.6$ eV or to the latent
heats $\Lambda =N_{A}\hbar \omega _{\max }\symbol{126}3.7\div 230$ kJ/mole.

These estimations seem rough ones, but non-contradictory.

It should be noted that in cases of narrow cooperation the indices of
correlations could accept, in accordance with (3.4), another values. This
problem must be investigated separately.

\qquad

{\large 5. EXPANSION OF TERMODYNAMIC POTENTIAL}

Let's consider the expansion of thermodynamic potential per unit volume in
terms of the order parameter $\eta $ (e.g. [14]) for homogeneous\ system as

$\Omega =\Omega _{0}+A\eta ^{2}+B\eta ^{4}-2\eta h$, \qquad \qquad (5.1)

and for nonhomogeneous system as

$\Omega =%
\displaystyle\int %
\{\Omega _{0}+A\eta ^{2}+B\eta ^{4}+g({\bf \nabla }h)^{2}-2\eta h\}dV$.
\qquad \qquad (5.1')

We shall begin with homogeneous systems. As the entropy close to transition
point is proportional to the temperature ''distance'' till critical point, $%
\Delta S\ \symbol{126}\ \epsilon $, then

$(\Omega -\Omega _{0})\ \symbol{126}\ \epsilon ^{2}\rightarrow \ 1/V_{c}$,
\qquad \qquad (5.2)

the dependence of energy of system on the density of interacting particles
seems natural and can be taken as the determinative one.

Minimization of (5.1) relative $\eta ^{2}$ at $h=0$ shows that $\eta
_{0}^{2}=-A/2B$. At $h\neq 0$ the condition of equilibrium $\partial \Omega
/\partial \eta =0$ leads to the equation: $A\eta +2B\eta ^{3}=h$, which
allows to determine the generalized susceptibility:

$\chi =\partial h/\partial \eta =1/(A+4B\eta ^{2})$. \qquad \qquad (5.3)

The conformity with (4.9) and (5.2) for the series coefficients and order
parameter require such representations:

$A(P,T)=a(P,T)\ R_{c}^{-2}=a(P)\ $sgn$(\epsilon )\ R_{c}^{-2}$,

$B(P,T)\ \symbol{126}\ b(P)\ R_{c}^{-1}$, \qquad \qquad \qquad (5.4)

$\eta \ \symbol{126}\ \eta _{0}\epsilon ^{\beta }$, $\qquad \beta =1/3$,
\qquad \qquad (5.5)

and therefore $A(P,T_{c})=0$.

With these coefficients

$\Omega =\Omega _{0}+a\ $sgn$(\epsilon )\ R_{c}^{-2}\ \eta ^{2}+b\
R_{c}^{-1}\ \eta ^{4}-2\eta h$, \qquad \qquad (5.1'')

and as from the dimension consideration $\eta h\ \symbol{126}\ \epsilon ^{2}$%
, it can be stated that

$h\ \symbol{126}\ \eta ^{d}$, $\qquad d=5$.\qquad \qquad\ (5.6)

Remind that in the Ginzburg-Landau theory [19] were initially proposed the
simplest representations: $A(P,T)=a(P)\ \epsilon $, $\ B(P,T)\ \symbol{126}\
B(P,T_{c})$. But experimental data require, in accordance at least with the
superconductivity results, their representation just in the forms leading to
(5.1'') [31].

The deduced indices are enough for the theory, others can be calculated via
the known interrelations. But their direct estimation may be of some
interest.

So, for the specific heat jumps from (5.2) follows that

$\Delta C_{p}\symbol{126}\ \partial ^{2}\Omega /\partial T^{2}\ \symbol{126}%
\ \epsilon ^{\alpha }$, $\qquad \alpha =0$, \qquad \qquad (5.7)

but the estimation (5.7) relates to the limiting, at $\hbar \omega
\rightarrow 0$, value of the scattering amplitude. Therefore $\alpha (\omega
)$ can be slightly distinct from zero at $\omega \neq 0$.

For the coefficient of surface tension, for example, from (5.2) follows the
estimation

$\sigma \ \symbol{126}\ \Omega (p,T)\ R_{\gamma }\ \symbol{126}\ \epsilon
^{2\ -\ \nu }$;\qquad \qquad\ (5.7')

the coefficient of thermal expansion can be written as

$\alpha _{P}\equiv (\partial \ln V_{c}/\partial T)_{P}=2\epsilon ^{-1}$
\qquad \qquad (5.7'')

and so on.

Let's consider some kinetic singularities. The simplest kinetic equation for
$\eta $ can be written as

$d\eta /dt=-\Gamma \ \partial \Omega /\partial \eta $. \qquad \qquad (5.8)

By the expansion of order parameter as $\eta =\eta _{0}+\delta \eta $ this
equation can be presented as

$d\delta \eta /dt=-\delta \eta /t_{0}$. \qquad \qquad (5.8')

In the linear approximation for $\delta \eta $ and with proposition $\eta
_{0}\ \symbol{126}\ A/2B$, we can estimate, in accordance with (4.9), that
the relaxation time would be expressed via

$1/t_{c,0}\ \symbol{126}\ 2A+12B\ \eta _{0}^{2}\rightarrow \ -8aR_{0}^{-2}\
\epsilon ^{4/3}\qquad \qquad $ (5.9)

(the Ginzburg-Landau theory leads to the linear dependence on $\epsilon $).

In the space nonhomogeneous system for this aim the expression (5.1') with
the order parameter $\eta (t,{\bf r})$ must be considered. Transition to the
partial Fourier transform $\eta (t,{\bf k})$ with taking into account the
space derivatives in (5.1') leads to the generalization of (5.9) as

$1/t_{c}({\bf k})\ \symbol{126}\ 1/t_{c,0}+a_{1}gk^{2}$.\qquad \qquad\ (5.10)

It means that the most intensive fluctuations take place near to $k_{0}%
\symbol{126}\ (a/g)^{1/2}\ \symbol{126}\ 1/R_{c}$. Its generalization in the
external field, $\eta \ \symbol{126}\ e^{-i(\omega t\ -\ {\bf kr})}$, leads
via the relation $\delta \eta ({\bf k})=\chi (\omega ,{\bf k})\ \eta (\omega
,{\bf k})$ to the determination of extended susceptibility:

$\chi (\omega ,{\bf k})\ \symbol{126}\ t_{c}(k)/[1-i\omega t_{c}(k)]$
.\qquad \qquad (5.11)

The expression (5.11) evidently generalizes the form (4.9) of susceptibility
of homogenous stationary substance and reduces to it at $\omega =k=0$.

The external field broadens transitions points till some zones, and as this
broadening can be estimated from the correspondence of the second and last
terms in (5.1), the relation will be slightly different from the usual one.

Let's compare the spontaneous and induced effects. As $\eta _{spont}^{2}\
\symbol{126}\ |a|/bR_{c}$ and $\eta _{ind}\ \symbol{126}\ \chi h$, the
inequality $\eta _{ind}$ $\geq $ $\eta _{spont}$ leads to the condition: $%
\eta _{T}$ $\geq $ $\epsilon ^{5/3}$, instead of the Landau condition: $\eta
_{T}\geq \epsilon ^{3/2}$. These functions allow also the estimations of
critical relaxation times by consideration of the fluctuation-type phenomena
at $kR_{c}>>1$ and the hydrodynamic or long-wave ones at $kR_{c}<<1$, etc.

\qquad

{\Large 5. ABOUT TWO-DIMENSIONAL SYSTEMS}

As the high temperature superconductors (cuprates, etc.), superfluid films
and so on represent layered structures, they combine properties of $3$-$D$
and $2$-$D$ substances. Therefore critical characteristics of such systems
may be intermediate between their limiting values.

Even for strictly two-dimensional systems amplitudes of $e$-$\gamma $
interactions would leave unchanged. But the optical theorem instead of (3.5)
is of such form [32]:

$\sigma _{tot}^{(2)}=(8\pi c/\omega )^{1/2}$\ Im $A(0)$. \qquad \qquad (6.1)

Together with (2.10) it leads to the ''volume'' (more correctly: the
''area'') of interaction:

$V_{\gamma }^{(2)}\symbol{126}\ \omega ^{-3/2}$ \qquad \qquad (6.2)

and the differences with $3$-$D$ case (3.6) are evident. Thus, if $V_{\gamma
}^{(2)}=4\pi (R_{\gamma }^{(2)})^{2}$, the radius of interaction for such
system $R_{\gamma }^{(2)}\ \symbol{126}\ \omega ^{-3/4}$, i.e. the critical
index is bigger than for $3$-$D$ systems (cf. [33]). It leads to such
estimation: for systems of thin layers the critical index for correlation
radius may be in the interval

$2/3<\nu _{2}<3/4$.\qquad \qquad (6.3)

More complete theory of pure $2$-$D$ system can be developed in the close
analogy with above considerations, but the layered systems require special
considerations.

\qquad

{\large CONCLUSIONS}

Such features of conducted researches would be emphasized.

1. The role of temporal functions in structure of basic thermodynamic
expressions is revealed and therefore the importance of temporal parameters
examination for problems of statistical mechanics is demonstrated.

2. The specific importance of investigation of particles/states transition
into the dressed state is elucidated.

3. The possibility of redirection of investigations of critical phenomena
onto properties of internal fields, i.e. the ways for transition from
phenomenology into field theory must be underlined.

4. The postulated condition of phase transition allows to examine the
simplest model of coexistence of some phases as corresponding levels,
separated by latent energy per particle, and the transitions between them as
a peculiar QED phenomenon. It conforms to the first principles of
microscopic theory, because ultimately reduces all considered processes to
the photon-electron interactions.

5. The correct system of critical indices, which has been deduced by more
formal procedures [1-3], is derived on physically evident level.

6. The developed theory explains the phenomenon of residual rays as the
manifestation of the phase-level model.

The considered model enables further refinements in different ways: by
consideration of complete propagators instead estimations of (2.10)-type and
more complete expressions of cross-sections, which will permit to consider
the dependence of transitions on parameters of systems and external factors.
More specific problems and some applications will be considered elsewhere.

\qquad

{\large APPENDIX}

Let us show, as example, that the conducted consideration allows to estimate
the relation between latent heat and temperature of gas condensation.

The bounce of free energy for transition at constant temperature can be
estimated via (2.5) and (2.8) by relation of delay durations in these phases:

$\Delta \Omega =\Omega _{k}-\Omega _{k-1}\approx -%
{\frac12}%
T\ \ln (\overline{\tau }_{k}/\overline{\tau }_{k-1})\ \symbol{126}-T\ \ln
(\Gamma _{k}/\omega _{0})$. \qquad \qquad (A.1)

In the infrared range, where characteristic emissions of latent heat may be
expected, the assessment $\omega _{0}/\Gamma _{k}\ \symbol{126}\ 10^{4}\div
10^{5}$ can be proposed, which leads for the latent heat $\Lambda
=N_{A}\Delta \Omega $ to an estimation

$\Lambda /T\approx R\ln (\Gamma _{k}/\omega _{0})\ \symbol{126}\ 18\div 23$
cal/mole, \qquad \qquad (A.2)

$R$ is the universal gas constant.

This estimation is very close to the well-known empirical Trawton rule,
established in 1884 for many substances (non-polar or weakly-polar): latent
heat and temperature of boiling are connected, at normal pressure, by the
relation $\Lambda _{b}/T_{b}\symbol{126}\ 21$ cal/mole K. This rule is often
used for physicochemical estimations (cf. [34], a number of its
specifications are continuously advancing, e.g. [35]). However, till now
this rule is not substantiated theoretically, its sense and significance
remain unknown. The estimation (A.2) can be indubitably concretized for
different transitions.

===============================

{\bf REFERENCES}

*). E-mail: mark\_perelman@mail.ru

[1]. H.E.Stanley. Rev.Mod.Phys., {\bf 71}, S358 (1999).

[2]. M.E.Fisher. Rev.Mod.Phys., {\bf 70}, 653 (1998).

[3]. G.A.Martynov. Phys. Uspekhi, {\bf 42}, 517 (1999).

[4]. V.Berestetski, E.Lifshitz, L.Pitayevskii. {\it Relativistic Quantum
Theory}. Pergamon, Oxford, 1971.

[5]. J.Sucher. Comments Modern Phys., {\bf 1}D, 227 (2000) and references
therein.

[6]. E.P.Wigner. Phys.Rev., {\bf 98}, 145 (1955), Wigner cited in this
connection the unpublished theses of L.Eisenbud.

[7]. F.T.Smith. Phys.Rev., {\bf 118}, 349; {\bf 119}, 2098 (1960); {\bf 130}%
, 394 (1963); J.Chem.Phys., {\bf 36}, 248; {\bf 38}, 1304 (1963).

[8]. M.L.Goldberger, K.M.Watson. {\it Collision Theory}. Wiley: NY, 1964.

[9]. E.Pollak, W.H.Miller. Phys.Rev.Lett., {\bf 53}, 115 (1984); E.Pollak.
J.Chem.Phys., {\bf 83}, 1111 (1986).

[10]. E.H.Hauge, J.A.St\o vneng. Rev.Mod.Phys. {\bf 61}, 917 (1989);
R.Landauer, Th.Martin. Rev.Mod.Phys. {\bf 66}, 217 (1994). See also the
reviews in: {\it Time in Quantum Mechanics }(J.G.Muga e.a., Ed's). Springer,
2002.

[11]. M.E.Perel'man. Phys.Lett. {\bf 32}A, 64 (1970); Sov.Phys. JETP, {\bf 31%
}, 1155 (1970) [Zh.Eksp.Teor.Fiz., {\bf 58}, 2139 (1970)]; M.E.Perel'man,
V.G.Arutyunian, Sov.Phys. JETP, {\bf 35}, 260 (1972) [Zh.Eksp.Teor.Fiz.,
{\bf 62}, 490 (1972)].

[12]. M.E.Perel'man. Sov.Phys. Doklady, {\bf 19}, 26 (1974) [DAN SSSR, {\bf %
214}, 539 (1974)].

[13]. G.E.Uhlenbeck, E.Beth. Physica, {\bf 3}, 729 (1936).

[14]. L.D.Landau, E.M.Lifshitz. {\it Statistical Physics}. I. (Any edition).

[15]. C.V.Heer. {\it Statistical Mechanics, Kinetic Theory and Stochastic
Processes}. Academic Press, 1972.

[16]. D.Bedeaux. Physica, {\bf 45}, 469; {\bf 46}, 17, 36 (1970) and
references therein.

[17]. R.Dashen, S.Ma, H.J.Bernstein. Phys.Rev, {\bf 187}, 345 (1969);
R.Dashen, S.Ma. J.Math.Phys., {\bf 11}, 1136 (1970).

[18]. M.E.Perel'man. {\it Preprints} arXiv: quant-ph/0311169;
Gen.Phys/0309123.

[19]. V.L.Ginzburg, L.D.Landau. Zh.Eksp.Teor.Fiz., {\bf 20}, 1064 (1950).

[20 ]. M.E.Perel'man. Phys.Lett.A, {\bf 32}, 411 (1971); Sov.Phys. Doklady,
{\bf 203},1030 (1972); Bull. Israel Phys.Soc., {\bf 48}, 10 (2002)..

[21]. M.E.Perel'man. Astrophysics, {\bf 17}, 383 (1981); In: {\it 2}$^{nd}$%
{\it \ Int. Sakharov Conference on Physics}, Moscow, May 1996. Singapore,
Int. Publ., 1997; M.E.Perel'man, I.Ya. Badinov. Bull.Acad.Sc. Georgian SSR,
{\bf 131}, 301 (1988); N.G.Khatiashvili, M.E.Perel'man. Phys. Earth \& Plan.
Inter., {\bf 57}, 169 (1989).

[22]. A.N.Mestvirishvili, I.G.Direktovich, S.I.Grigoriev, M.E.Perel'man.
Phys.Lett.A, {\bf 60}, 143 (1977); V.A.Tatarchenko. Crystallography, {\bf 24}%
, 408 (1979); V.A.Tatarchenko, L.M.Umarov. Ibid., {\bf 25}, 1311 (1980);
L.M.Umarov, V.A.Tatarchenko. Ibid., {\bf 29}, 1146 (1984); M.I.Molotsky,
B.P.Peregud. Sov.Tech.Phys., {\bf 51}, 618 (1981); K.B.Abramova et al. Sov.
Opt. Spectr., {\bf 58}, 809 (1985).

[23]. M.E.Perel'man, G.M.Rubinstein. Sov.Phys. Doklady, {\bf 17}, 352 (1972)
[DAN SSSR, {\bf 203}, 798 (1972)]. Further researches are summarized in:
M.E.Perel'man. {\it Kinetical Quantum Theory of Optical Dispersion}.
Tbilisi: Mezniereba, 1989, 140 p. (In Russian)

[24]. F.W.J.Olver. {\it Asymptotics and Special Functions}. NY, Academic
Press: 1974 (Ch.14-3).

[25 ]. S.Ma. {\it Modern Theories of Critical Phenomena}. W.A.Benjamin, L.,
1976.

[26]. A.Z.Patachinsky, V.L.Pokrovsky. {\it Fluctuation Theory of Phase
transitions}. Pergamon, 1979.

[27]. E.M.Lifshitz, L.P.Pitaevskii. {\it Statistical Physic}s. II. Pergamon

[28]. E.C.Baly. Phil.Mag., {\bf 47}, 15 (1920); Cl.Schaeffer, F.Matossi.
{\it Das Iltrarote Spektrum}. B., J.Springer, 1930.

[29]. C.Kittel. {\it Quantum Theory of Solids}. NY, Wiley, 1963.

[30]. L.H.Ryder. {\it Quantum Field Theory}. Cambridge Univ. Press, 1985.

[31]. V.L.Ginzburg. Physics-Uspekhi, {\bf 40}, 407 (1997); {\bf 43}, 573
(2000) [UFN, {\bf 167}, 429 (1997); {\bf 170}, 619 (2001)] and references
therein.

[32]. L.D.Landau, E.M.Lifshitz. {\it Quantum Mechanics}. 2$^{nd}$ ed. NY:
Pergamon, 1965.

[33]. H.B.Tarko, M.E.Fisher. Phys.Rev.B, {\bf 11}, 1217 (1975).

[34]. R.M.White, Th.H.Geballe. {\it Long Range Order in Solids}. Academic
Press, 1979.

[35]. K.Shimoda. J.Chem.Phys., {\bf 78}, 4784, 1983; S.R.Logan. Z.Naturfor.,
{\bf 53}A, 178, 1998; V.V.Ovchinikov et al. J.Mol.Liquids, {\bf 91}, 47,
2001.

\end{document}